\documentclass[12pt,twoside,dvips]{article}
\usepackage{amssymb}
\usepackage{amsfonts}
\usepackage{amsmath}
\usepackage{pifont}
\usepackage[spanish,english]{babel}
\usepackage{graphicx}
\usepackage{verbatim}

\begin{document}
\title{Radiative seesaw-type mechanism of quark masses in $%
SU\left(3\right)_{C}\otimes SU\left(3\right)_{L}\otimes U\left(1\right)_{X}$}
\author{A. E. C\'arcamo Hern\'andez$^{1}\thanks{e-mail: antonio.carcamo@usm.cl}$,
R. Mart\'{\i}nez$^{2}\thanks{e-mail: remartinezm@unal.edu.co}$,
F. Ochoa$^{2}\thanks{e-mail: faochoap@unal.edu.co}$
\and $^{1}$\textit {Universidad T\'{e}cnica Federico Santa Mar\'{\i}a}\\
\textit{and Centro Cient\'{\i}fico-Tecnol\'{o}gico de Valpara\'{\i}so}\\
\textit{Casilla 110-V, Valpara\'{\i}so, Chile,}\\ \and
$^{2}$\textit{Departamento de F\'{\i}sica, Universidad Nacional de Colombia,} \\
\textit{Ciudad Universitaria, Bogot\'{a} D.C., Colombia. }}
\date{\today }

\maketitle

\begin{abstract}
We take up again the study of the mass spectrum of the quark sector in a
model with gauge symmetry $SU(3)_{c}\otimes SU(3)_{L}\otimes U(1)_{X}$ (331).  In a special type II-like 331 model, we obtain specific zero-texture mass matrices for the quarks which predict four massless quarks ($u,c,d,s$) and two massive quarks
($b,t$) at the electroweak scale ($\sim $ GeV).  By considering the mixing between the
SM quarks and new exotic quarks at large scales predicted by the
model, we find that a third quark (associated to the charm quark) acquires a mass. The remaining
light quarks ($u,d,s$) get small masses ($\sim$ MeV) via radiative corrections. 
\end{abstract}

\maketitle



\section{Introduction}

The ATLAS and CMS experiments at the CERN Large Hadron Collider (LHC) have
found a $126$ GeV Higgs boson \cite%
{atlashiggs,cmshiggs,newtevatron,CMS-PAS-HIG-12-020} through the $%
h\rightarrow \gamma \gamma $ decay channel, increasing our knowledge of the
Electroweak Symmetry Breaking (EWSB) sector and opening a new era in
particle physics. Now the priority of the LHC experiments will be to measure
precisely the couplings of the new particle to standard model fermions and
gauge bosons and to establish its quantum numbers. It also remains to look
for further associated with the EWSB mechanism which will allow to
discriminate among the different theoretical models addressed to explain
EWSB.

Despite all its success, the standard model (SM) of the electroweak
interactions based on the $SU(3)_{C}\otimes SU(2)_{L}\otimes U(1)_{Y}$ gauge
symmetry has many unexplained features \cite{SM}. Most of them are linked to
the mechanism responsible for the stabilization of the weak scale, the
origin of fermion masses and mixings and the three family structure. Because
of this reason, many people consider the Standard Model to be an effective
framework of a yet unknown more fundamental theory. A fundamental theory,
one expects, should have a dynamical explanation for the masses and mixings.
The lack of predictivity of the fermion masses and mixings in the SM has
motivated many models based on extended symmetries in the context of Two
Higgs Doublets, Grand Unification, Extradimensions and Superstrings, leading
to specific textures for the Yukawa couplings \cite{GUT,Extradim,String}.
The understanding of the discrete flavor symmetries hidden in such textures
may be useful in the knowledge of the underlying dynamics responsible for
quark mass generation and CP violation. One clear and outstanding feature in
the pattern of quark masses is that they increase from one generation to the
next spreading over a range of five orders of magnitude, and that the
mixings from the first to the second and to the third family are in
decreasing order \cite{PDG,Fritzsch,Fx,Xing2011}. From the phenomenological
point of view, it is possible to describe some features of the mass
hierarchy by assuming zero-texture Yukawa matrices \cite{textures}. Models
with spontaneously broken flavor symmetries may also produce hierarchical
mass structures. These horizontal symmetries can be continuous and Abelian,
as the original Froggatt-Nielsen model \cite{froggatt}, or non-Abelian as
for example SU(3) and SO(3) family models \cite{non-abelian}. Models with
discrete symmetries may also predict mass hierarchies for leptons \cite%
{discrete-lepton} and quarks \cite{discrete-quark}. Other models with
horizontal symmetries have been proposed in the literature \cite{horizontal}.

On the other hand, the origin of the family structure of the fermions can be
addressed in family dependent models where a symmetry distinguish fermions
of different families. Alternatively, an explanation to this issue can also
be provided by the models based on the gauge symmetry $SU(3)_{c}\otimes
SU(3)_{L}\otimes U(1)_{X}$, also called 3-3-1 models, which introduce a
family non-universal $U(1)_{X}$ symmetry \cite{331-pisano, 331-frampton,
331-long, M-O}. These models have a number of phenomenological advantages.
First of all, the three family structure in the fermion sector can be
understood in the 3-3-1 models\ from the cancellation of chiral anomalies 
\cite{anomalias} and asymptotic freedom in QCD. Secondly, the fact that the
third family is treated under a different representation, can explain the
large mass difference between the heaviest quark family and the two lighter
ones \cite{third-family}. Finally, these models contain a natural
Peccei-Quinn symmetry, necessary to solve the strong-CP problem \cite{PC}.

The 3-3-1 models extend the scalar sector of the SM into three $SU(3)_{L}$
scalar triplets: one heavy triplet field with a Vacuum Expectation Value
(VEV) at high energy scale $\langle \chi \rangle =\nu _{\chi }$, which
breaks the symmetry $SU(3)_{L}\otimes U(1)_{X}$ into the SM electroweak
group $SU(2)_{L}\otimes U(1)_{Y}$, and two lighter triplets with VEVs at the
electroweak scale $\langle \rho \rangle =\upsilon _{\rho }$ and $\langle
\eta \rangle =\upsilon _{\eta }$, which trigger Electroweak Symmetry
Breaking. Besides that, the 3-3-1 model could possibly explain
the excess of events in the $h\rightarrow \gamma \gamma $ decay, recently
observed at the LHC, since the heavy exotic quarks, the charged Higges and
the heavy charged gauge bosons contribute to this process. On the other
hand, the 3-3-1 model reproduces an specialized Two Higgs Doublet Model type
III (2HDM-III) in the low energy limit, where both triplets $%
\rho $ and $\eta $ are decomposed into two hypercharge-one $SU(2)_{L}$
doublets plus charged and neutral singlets. 
Thus, like the 2HDM-III, the 3-3-1 model can predict huge flavor changing
neutral currents (FCNC) and CP-violating effects, which are severely
suppressed by experimental data at electroweak scales. In the 2HDM-III, for
each quark type, up or down, there are two Yukawa couplings. One of the
Yukawa couplings is for generating the quark masses, and the other one
produces the flavor changing couplings at tree level. One way to remove both the
huge FCNC and CP-violating effects, is by
imposing discrete symmetries, obtaining two types of 3-3-1 models (type I
and II models), which exhibit the same Yukawa interactions as the 2HDM type
I and II at low energy where each fermion is coupled at most to one Higgs
doublet. In the 3-3-1 model type I, one Higgs electroweak triplet (for
example, $\rho $) provide masses to the phenomenological up- and down-type
quarks, simultaneously. In the type II, one Higgs triplet ($\eta $) gives
masses to the up-type quarks and the other triplet ($\rho $) to the
down-type quarks. Recently, authors in ref. \cite{331-2hdm} discuss the mass structures in the framework of the I-type 331 model. 
In this paper we obtain different structures for the type II-like model. We found that only the top and bottom
quarks acquire masses if the mixing of the SM quarks with
the exotic quarks is neglected. We obtain by the method of recursive
expansion \cite{grimus} that if mixing couplings with the heavy quark sector
of the 3-3-1 model are considered, only the charm quark obtains a mass, while the light quarks remains massless.  The masses of the up, down and strange quarks are generated through loop
corrections which is a kind of seesaw-like radiative mechanism that involves
the virtual exotic quarks as well as neutral and charged scalars running in
the loops. Thus, the hierarchy of the quark mass spectrum can be explained from three different sources:
the tree level quark mass matrices from the symmetry breaking, the mixings between the SM quarks
and the exotic quarks, and seesaw-like radiative corrections.
This mechanism of generating the quark masses provides an alternative to
that ones discussed in Refs \cite{effective}, \cite{effective-331} where
effective operators and one-loop corrections are introduced.

This paper is organized as follows. In Section 2 we briefly describe some
theoretical aspects of the 3-3-1 model and its particle content, in
particular in the fermionic and scalar sector in order to obtain the mass
spectrum. Section 3 is devoted to discuss possible zero-textures for the SM quark mass matrices at tree
level. In section 4 we obtain the quark masses at tree and one-loop level of the complete model by imposing an specific zero-texture masses. Finally in Sec. 5, we state our conclusions.

\section{The fermion and scalar sector}

We consider the 3-3-1 model where the electric charge is defined by:

\begin{equation}
Q=T_{3}-\frac{1}{\sqrt{3}}T_{8}+X,
\end{equation}%
with $T_{3}=\frac{1}{2}Diag(1,-1,0)$ and $T_{8}=(\frac{1}{2\sqrt{3}}%
)Diag(1,1,-2)$. In order to avoid chiral anomalies, the model introduces in
the fermionic sector the following $(SU(3)_{c},SU(3)_{L},U(1)_{X})$
left-handed representations: 
\begin{eqnarray}
Q_{L}^{1} &=&%
\begin{pmatrix}
U^{1} \\ 
D^{1} \\ 
T^{1} \\ 
\end{pmatrix}%
_{L}:(3,3,1/3),\left\{ 
\begin{array}{c}
U_{R}^{1}:(3^{\ast },1,2/3) \\ 
D_{R}^{1}:(3^{\ast },1,-1/3) \\ 
T_{R}^{1}:(3^{\ast },1,2/3) \\ 
\end{array}%
\right.  \notag \\
Q_{L}^{2,3} &=&%
\begin{pmatrix}
D^{2,3} \\ 
U^{2,3} \\ 
J^{2,3} \\ 
\end{pmatrix}%
_{L}:(3,3^{\ast },0),\left\{ 
\begin{array}{c}
D_{R}^{2,3}:(3^{\ast },-1/3) \\ 
U_{R}^{2,3}:(3^{\ast },1,2/3) \\ 
J_{R}^{2,3}:(3^{\ast },1,-1/3) \\ 
\end{array}%
\right.  \notag \\
L_{L}^{1,2,3} &=&%
\begin{pmatrix}
\nu ^{1,2,3} \\ 
e^{1,2,3} \\ 
(\nu ^{1,2,3})^{c} \\ 
\end{pmatrix}%
_{L}:(1,3,-1/3),\left\{ 
\begin{array}{c}
e_{R}^{1,2,3}:(1,1,-1) \\ 
N_{R}^{1,2,3}:(1,1,0) \\ 
\end{array}%
\right.  \label{fermion_spectrum}
\end{eqnarray}%
where $U_{L}^{i}$ and $D_{L}^{i}$ for $i=1,2,3$ are three up- and down-type
quark components in the flavor basis, while $\nu _{L}^{i}$ and $e_{L}^{i}$
are the neutral and charged lepton families. The right-handed sector
transform as singlets under $SU(3)_{L}$ with $U(1)_{X}$ quantum numbers
equal to the electric charges. In addition, we see that the model introduces
heavy fermions with the following properties: a single flavor quark $T^{1}$
with electric charge $2/3$, two flavor quarks $J^{2,3}$ with charge $-1/3$,
three neutral Majorana leptons $(\nu ^{1,2,3})_{L}^{c}$ and three
right-handed Majorana leptons $N_{R}^{1,2,3}$. 
On the other hand, the scalar sector introduces one triplet
field with VEV $\langle \chi \rangle _{0}=\upsilon _{\chi }$, which provides
the masses to the new heavy fermions, and two triplets with VEVs $\langle
\rho \rangle _{0}=\upsilon _{\rho }$ and $\langle \eta \rangle _{0}=\upsilon
_{\eta }$, which give masses to the SM-fermions at the electroweak scale. However, as it will be shown in section 4, we can have a discrete symmetry in the quark sector that allows the triplet
$\chi$ to give masses not only to the heavy exotic quarks but also to the light quarks via radiative see-saw like mechanism while the triplets $\rho$ and $\eta$ give masses to the remaining quarks.
The $(SU(3)_{L},U(1)_{X})$ group structure of the scalar fields are:

\begin{eqnarray}
\chi&=& 
\begin{pmatrix}
\chi_1^{0} \\ 
\chi_2^{-} \\ 
\frac{1}{\sqrt{2}}(\upsilon_{\chi} + \xi_{\chi} \pm i \zeta_{\chi} ) \\ 
\end{pmatrix}%
: (3,-1/3)  \notag \\
\rho&=& 
\begin{pmatrix}
\rho_1^{+} \\ 
\frac{1}{\sqrt{2}}(\upsilon_{\rho} + \xi_{\rho} \pm i \zeta_{\rho} ) \\ 
\rho _3^{+} \\ 
\end{pmatrix}
: (3,2/3)  \notag \\
\eta &=& 
\begin{pmatrix}
\frac{1}{\sqrt{2}}(\upsilon_{\eta} + \xi_{\eta} \pm i \zeta_{\eta} ) \\ 
\eta _2^{-} \\ 
\eta _3^{0}%
\end{pmatrix}%
:(3,-1/3).  \label{331-scalar}
\end{eqnarray}

The EWSB follows the scheme ${SU(3)_L\otimes U(1)_X} {%
\xrightarrow{\langle
\chi \rangle}} {SU(2)_L\otimes U(1)_Y} {%
\xrightarrow{\langle \eta
\rangle,\langle \rho \rangle}} {U(1)_Q}, $ where the vacuum expectation
values satisfy $v_{\chi }\gg v_{\eta },v_{\rho }.$

The interactions among the scalar fields are contained in the following most
general potential that we can construct with three scalar triplets: 
\begin{eqnarray}
&&V_{H}=\mu _{\chi }^{2}(\chi ^{\dagger }\chi )+\mu _{\eta }^{2}(\eta
^{\dagger }\eta )+\mu _{\rho }^{2}(\rho ^{\dagger }\rho )+f\left( \chi
_{i}\eta _{j}\rho _{k}\varepsilon ^{ijk}+H.c.\right) +\lambda _{1}(\chi
^{\dagger }\chi )(\chi ^{\dagger }\chi )  \notag \\
&&+\lambda _{2}(\rho ^{\dagger }\rho )(\rho ^{\dagger }\rho )+\lambda
_{3}(\eta ^{\dagger }\eta )(\eta ^{\dagger }\eta )+\lambda _{4}(\chi
^{\dagger }\chi )(\rho ^{\dagger }\rho )+\lambda _{5}(\chi ^{\dagger }\chi
)(\eta ^{\dagger }\eta )  \notag \\
&&+\lambda _{6}(\rho ^{\dagger }\rho )(\eta ^{\dagger }\eta )+\lambda
_{7}(\chi ^{\dagger }\eta )(\eta ^{\dagger }\chi )+\lambda _{8}(\chi
^{\dagger }\rho )(\rho ^{\dagger }\chi )+\lambda _{9}(\rho ^{\dagger }\eta
)(\eta ^{\dagger }\rho ).  \label{v00}
\end{eqnarray}

After the symmetry breaking, it is found that the mass eigenstates are
related to the weak states in the scalar sector by \cite{331-long, M-O}:

\begin{eqnarray}
\begin{pmatrix}
G_{1}^{\pm } \\ 
H_{1}^{\pm } \\ 
\end{pmatrix}%
=R_{\beta _{T}}%
\begin{pmatrix}
\rho _{1}^{\pm } \\ 
\eta _{2}^{\pm } \\ 
\end{pmatrix}
&,&\hspace{0.3cm}%
\begin{pmatrix}
G_{1}^{0} \\ 
A_{1}^{0} \\ 
\end{pmatrix}%
=R_{\beta _{T}}%
\begin{pmatrix}
\zeta _{\rho } \\ 
\zeta _{\eta } \\ 
\end{pmatrix}%
,\hspace{0.3cm}%
\begin{pmatrix}
H_{1}^{0} \\ 
h^{0} \\ 
\end{pmatrix}%
=R_{\alpha _{T}}%
\begin{pmatrix}
\xi _{\rho } \\ 
\xi _{\eta } \\ 
\end{pmatrix}
\label{331-mass-scalar-a} \\
\begin{pmatrix}
G_{2}^{0} \\ 
H_{2}^{0} \\ 
\end{pmatrix}%
=R%
\begin{pmatrix}
\chi _{1}^{0} \\ 
\eta _{3}^{0} \\ 
\end{pmatrix}
&,&\hspace{0.3cm}%
\begin{pmatrix}
G_{3}^{0} \\ 
H_{3}^{0} \\ 
\end{pmatrix}%
=\frac{-1}{\sqrt{2}}R%
\begin{pmatrix}
\zeta _{\chi } \\ 
\xi _{\chi } \\ 
\end{pmatrix}%
,\hspace{0.3cm}%
\begin{pmatrix}
G_{2}^{\pm } \\ 
H_{2}^{\pm } \\ 
\end{pmatrix}%
=R%
\begin{pmatrix}
\chi _{2}^{\pm } \\ 
\rho _{3}^{\pm } \\ 
\end{pmatrix}%
,  \label{331-mass-scalar-b}
\end{eqnarray}

with

\begin{equation}
R_{\alpha _{T}(\beta_{T})}=\left( 
\begin{array}{cc}
\cos \alpha _{T} (\beta_{T}) & \sin \alpha _{T}(\beta_{T}) \\ 
-\sin \alpha _{T}(\beta_{T}) & \cos \alpha _{T}(\beta_{T})%
\end{array}%
\right) ,
\hspace{2cm}R=\left( 
\begin{array}{cc}
-1 & 0 \\ 
0 & 1%
\end{array}%
\right)
\end{equation}

where $\tan \beta _{T}=v_{\eta }/v_{\rho }$, and $\tan 2\alpha _{T}=M_1/(M_2-M_3)$ with:

\begin{eqnarray}
M_1&=&4\lambda _6 v_{\eta}v_{\rho}+2\sqrt{2}fv_{\chi}, \nonumber \\
M_2&=&4\lambda _2 v_{\rho}^2-\sqrt{2}fv_{\chi}\tan \beta _T , \nonumber \\
M_3&=&4\lambda _3 v_{\eta}^2-\sqrt{2}fv_{\chi}/\tan \beta _T .
\end{eqnarray}

With the above spectrum, we obtain the following $SU(3)_{L}\otimes U(1)_{X}$
renormalizable Yukawa Lagrangian for the quark sector \cite{331-2hdm}: 

\begin{eqnarray}
-\mathcal{L}_{Y} &=&\overline{Q}_{L}^{1}\left( \eta h_{\eta 1j}^{U}+\chi
h_{\chi 1j}^{U}\right) U_{R}^{j}+\overline{Q}_{L}^{1}\rho h_{\rho
1j}^{D}D_{R}^{j}  \notag \\
&+&\overline{Q}_{L}^{1}\rho h_{\rho 1m}^{J}J_{R}^{m}+\overline{Q}%
_{L}^{1}\left( \eta h_{\eta 11}^{T}+\chi h_{\chi 11}^{T}\right) T_{R}^{1} 
\notag \\
&+&\overline{Q}_{L}^{n}\rho ^{\ast }h_{\rho nj}^{U}U_{R}^{j}+\overline{Q}%
_{L}^{n}\left( \eta ^{\ast }h_{\eta nj}^{D}+\chi ^{\ast }h_{\chi
nj}^{D}\right) D_{R}^{j}  \notag \\
&+&\overline{Q}_{L}^{n}\left( \eta ^{\ast }h_{\eta nm}^{J}+\chi ^{\ast
}h_{\chi nm}^{J}\right) J_{R}^{m}+\overline{Q}_{L}^{n}\rho ^{\ast }h_{\rho
n1}^{T}T_{R}^{1}+h.c,  \label{331-yukawa}
\end{eqnarray}%
where $n=2,3$ is the index that label the second and third quark triplet
shown in Eq. (\ref{fermion_spectrum}), and $h_{\phi ij}^{f}$ are the ${i,j}$
components of non-diagonal matrices in the flavor space associated with each
scalar triplet $\phi :\eta ,\rho ,\chi $. 

\section{Zero-texture masses at low energy}

By considering a scenario where the mixing terms among fields at small and large mass scales in Eq. (\ref{331-yukawa}) do not contribute at low energy, we obtain the following
decoupled low energy Yukawa Lagrangian:

\begin{eqnarray}
-\mathcal{L}_{Y}^{LE} &=&\overline{Q}_{L}^{1}\eta h_{\eta
1j}^{U}U_{R}^{j}+\overline{Q}_{L}^{n}\eta ^{\ast }h_{\eta nj}^{D}D_{R}^{j} 
\notag \\
&&+\overline{Q}_{L}^{n}\rho ^{\ast }h_{\rho nj}^{U}U_{R}^{j}+\overline{Q}%
_{L}^{1}\rho h_{\rho 1j}^{D}D_{R}^{j}+h.c.  \notag
\end{eqnarray}

Although the above Lagrangian exhibits the same general form as the 2HDM Lagrangian, they are not the same because the abelian sector $U(1)_X$ of the $3-3-1$ symmetry introduces a quantum
number that differentiates the second and third rows from the first one, from which not all Yukawa couplings are allowed by the symmetry. In
2HDM models there are no labels and the three rows are identicals. Thus, by imposing appropriate discrete symmetries, many ansatze for the couplings can be obtained. From the previous expression
it follows that the mass Lagrangian corresponding to the SM quark sector
is given by:

\begin{eqnarray}
-\mathcal{L}_{Y}^{mass-SM} &=&\frac{v_{\eta }}{\sqrt{2}}\overline{U}%
_{L}^{1}h_{\eta 1j}^{U}U_{R}^{j}+\frac{v_{\eta }}{\sqrt{2}}\overline{D}%
_{L}^{n}h_{\eta nj}^{D}D_{R}^{j}  \notag \\
&+&\frac{v_{\rho }}{\sqrt{2}}\overline{U}_{L}^{n}h_{\rho nj}^{U}U_{R}^{j}+%
\frac{v_{\rho }}{\sqrt{2}}\overline{D}_{L}^{1}h_{\rho 1j}^{D}D_{R}^{j}+h.c, 
\end{eqnarray}

The 3-3-1 model gives the different possible textures
that can be chosen independently for the up and down sector according to the
discrete symmetry imposed. These textures are given by:


\begin{equation}
M_{U}^{(A)}=\frac{v_{\eta }}{\sqrt{2}}\left( 
\begin{array}{cc}
h_{\eta 11}^{U} & h_{\eta 1m}^{U} \\ 
0_{2\times 1} & 0_{2\times 2}%
\end{array}%
\right) ,\hspace{1cm}M_{U}^{(B)}=\frac{v_{\rho }}{\sqrt{2}}\left( 
\begin{array}{cc}
0 & 0_{1\times 2} \\ 
h_{\rho n1}^{U} & h_{\rho nm}^{U}%
\end{array}%
\right) ,
\label{U-textures}
\end{equation}

\begin{equation}
M_{D}^{(A)}=\frac{v_{\eta }}{\sqrt{2}}\left( 
\begin{array}{cc}
0 & 0_{1\times 2} \\ 
h_{\eta n1}^{D} & h_{\eta nm}^{D}%
\end{array}%
\right) ,\hspace{1cm}M_{D}^{(B)}=\frac{v_{\rho }}{\sqrt{2}}\left( 
\begin{array}{cc}
h_{\rho 11}^{D} & h_{\rho 1m}^{D} \\ 
0_{2\times 1} & 0_{2\times 2}%
\end{array}%
\right)
\label{D-textures} ,
\end{equation}
with $n$ and $m$ $=2,3$. The choice of the textures $M_{U}^{(A)}$ and $M_{D}^{(B)}$ (where the subindices $U$
and $D$ refer to the up and down sector, respectively) can be obtained by imposing:

\begin{eqnarray}
U_R \rightarrow -U_R, \hspace {1cm} D_R \rightarrow D_R, \hspace {1cm}  \rho \rightarrow \rho, \hspace {1cm} \eta \rightarrow -\eta.
\label{discrete_1} 
\end{eqnarray}
These textures implies that the quarks generates mass from the $Q_{L}^{1}$ terms at tree level. In this case only the
top and bottom quarks would acquire mass while the other quarks will remain
massless. The textures $M_{U}^{(B)}$ and $M_{D}^{(A)}$ can be obtained through:

\begin{eqnarray}
U_R \rightarrow -U_R, \hspace {1cm} D_R \rightarrow D_R, \hspace {1cm}  \rho \rightarrow - \rho, \hspace {1cm} \eta \rightarrow \eta.
\label{discrete_2} 
\end{eqnarray}
These would imply $4$ massive quarks from the $Q_L^n$ terms
at tree level, which will be unnatural since the masses will be practically
generated by hand through the Yukawa couplings. 
The choice of $M_{U}^{(A)}$ and $M_{D}^{(A)}$, where

\begin{eqnarray}
U_R \rightarrow -U_R, \hspace {1cm} D_R \rightarrow -D_R, \hspace {1cm}  \rho \rightarrow \rho, \hspace {1cm} \eta \rightarrow -\eta
\label{discrete_3} 
\end{eqnarray}
will generate a mass for the top quark, which comes from the $Q_{L}^{1}$ term, while the
bottom quark coming from the $Q_{L}^{n}$ terms will be massless and the $%
d$ and $s$ quarks will be massive at tree level. Thus, the
choice of this textures does not lead to a phenomenological
viable quark mass spectrum. The choice of $M_{U}^{(B)}$ and $M_{D}^{(B)}$ with

\begin{eqnarray}
U_R \rightarrow -U_R, \hspace {1cm} D_R \rightarrow -D_R, \hspace {1cm}  \rho \rightarrow -\rho, \hspace {1cm} \eta \rightarrow \eta
\label{discrete_4} 
\end{eqnarray}
would imply that only three quarks will
be massive. In the up sector, one can choose the massive top and charm
quarks as elements of $Q_{L}^{n}$. In that case, the bottom and strange will
be massless while the down quark coming from $Q_{L}^{1}$ will acquire mass,
which is unnatural. In conclusion, the textures $M_{U}^{(A)}$ and $M_{D}^{(B)}$ with the discrete symmetry in Eqn. (\ref{discrete_1}) could provide a
better explanation for the quark mass hierarchy. The vanishing entries
will be filled by the mixings between the SM and exotic quarks or by
radiative corrections. 

\section{Zero-texture masses with mixing couplings}

In order to obtain the submatrices $M_{U}^{(A)}$ and $M_{D}^{(B)}$ in Eqs. (\ref{U-textures}) and (\ref{D-textures}) from the original 331 Lagrangian in (\ref{331-yukawa}),  we extend the discrete symmetry in (\ref{discrete_1}) to:  
\begin{eqnarray}
U_{R}\rightarrow -U_{R}&,&\hspace{1cm} D_{R}\rightarrow D_{R}, \hspace{1cm} \eta \rightarrow -\eta,\hspace{1cm} \rho \rightarrow \rho, %
 \nonumber \\
&\chi& \rightarrow \chi,\hspace{1cm}%
T_{R}\rightarrow T_{R},\hspace{1cm}J_{R}\rightarrow J_{R}
\end{eqnarray}%
which restricts the $SU\left( 3\right) _{L}\times U\left( 1\right) _{X}$
renormalizable Yukawa Lagrangian to be given by: 
\begin{eqnarray}
-\mathcal{L}_{Y} &=&\overline{Q}_{L}^{1}\eta h_{\eta 1j}^{U}U_{R}^{j}+\overline{Q}_{L}^{1}\rho h_{\rho 1j}^{D}D_{R}^{j}%
 \notag \\
&+&\overline{Q}_{L}^{1}\chi h_{\chi 11}^{T}T_{R}^{1}+\overline{Q}_{L}^{n}\chi ^{\ast }h_{\chi
nm}^{J}J_{R}^{m} \notag \\
&+&\overline{Q}_{L}^{n}\rho^{\ast }h_{\rho n1}^{T}T_{R}^{1} +\overline{Q}%
_{L}^{n}\chi ^{\ast }h_{\chi nj}^{D}D_{R}^{j}+\overline{Q}_{L}^{1}\rho
h_{\rho 1m}^{J}J_{R}^{m}+h.c.
\label{yukawaAB}
\end{eqnarray}%


The previous Lagrangian can be rewritten as: 
\begin{equation}
-\mathcal{L}_{Y}=-\mathcal{L}_{Y}^{\left( 1\right) }-\mathcal{L}_{Y}^{\left(
2\right) }-\mathcal{L}_{Y}^{\left( 3\right) }
\end{equation}%
where $-\mathcal{L}_{Y}^{\left( 1\right) }$ correspond to the quark mass
Lagrangian, while $-\mathcal{L}_{Y}^{\left( 2\right) }$ and $-\mathcal{L}%
_{Y}^{\left( 3\right) }$ are the Lagrangians which include the interactions
of the quarks with the neutral and charged Higgs and Goldstone bosons, respectively. These
Lagrangians are given by: 
\begin{eqnarray}
-\mathcal{L}_{Y}^{\left( 1\right) } &=&\frac{v_{\eta }}{\sqrt{2}}\overline{U}%
_{L}^{1}h_{\eta 1j}^{U}U_{R}^{j}+\frac{v_{\rho }}{\sqrt{2}}\overline{U}%
_{L}^{n}h_{\rho n1}^{T}T_{R}^{1}+\frac{v_{\chi }}{\sqrt{2}}\overline{T}%
_{L}^{1}h_{\chi 11}^{T}T_{R}^{1} +\frac{v_{\rho }}{\sqrt{2}}\overline{D}_{L}^{1}h_{\rho 1j}^{D}D_{R}^{j} \notag \\
&&+%
\frac{v_{\rho }}{\sqrt{2}}\overline{D}_{L}^{1}h_{\rho 1m}^{J}J_{R}^{m}+\frac{%
v_{\chi }}{\sqrt{2}}\overline{J}_{L}^{n}h_{\chi nj}^{D}D_{R}^{j}+\frac{%
v_{\chi }}{\sqrt{2}}\overline{J}_{L}^{n}h_{\chi nm}^{J}J_{R}^{m}+h.c 
\label{lqm}
\end{eqnarray}%
\begin{eqnarray}
-\mathcal{L}_{Y}^{\left( 2\right) } &=&\frac{1}{\sqrt{2}}\overline{U}%
_{L}^{1}\left( \xi _{\eta }+i\zeta _{\eta }\right) h_{\eta 1j}^{U}U_{R}^{j}+%
\overline{U}_{L}^{1}\chi _{1}^{0}h_{\chi 11}^{T}T_{R}^{1}+\overline{T}%
_{L}^{1}\eta _{3}^{0}h_{\eta 1j}^{U}U_{R}^{j}  \notag \\
&&+\frac{1}{\sqrt{2}}\overline{T}_{L}^{1}\left( \xi _{\chi }+i\zeta _{\chi
}\right) h_{\chi 11}^{T}T_{R}^{1}+\frac{1}{\sqrt{2}}\overline{U}%
_{L}^{n}\left( \xi _{\rho }-i\zeta _{\rho }\right) h_{\rho n1}^{T}T_{R}^{1} 
\notag \\
&&+\frac{1}{\sqrt{2}}\overline{D}_{L}^{1}\left( \xi _{\rho }+i\zeta _{\rho
}\right) h_{\rho 1j}^{D}D_{R}^{j}+\frac{1}{\sqrt{2}}\overline{D}%
_{L}^{1}\left( \xi _{\rho }+i\zeta _{\rho }\right) h_{\rho 1m}^{J}J_{R}^{m}+%
\overline{D}_{L}^{n}\chi _{1}^{0}h_{\chi nj}^{D}D_{R}^{j}  \notag \\
&&+\overline{D}_{L}^{n}\chi _{1}^{0}h_{\chi nm}^{J}J_{R}^{m}+\frac{1}{\sqrt{2%
}}\overline{J}_{L}^{n}\left( \xi _{\chi }-i\zeta _{\chi }\right) h_{\chi
nj}^{D}D_{R}^{j}+\frac{1}{\sqrt{2}}\overline{J}_{L}^{n}\left( \xi _{\chi
}-i\zeta _{\chi }\right) h_{\chi nm}^{J}J_{R}^{m}+h.c \notag \\
\end{eqnarray}%
\begin{eqnarray}
&-\mathcal{L}_{Y}^{\left( 3\right) }=&+\overline{D}_{L}^{n}\rho
_{1}^{-}h_{\rho n1}^{T}T_{R}^{1}+\overline{J}_{L}^{n}\rho _{3}^{-}h_{\rho
n1}^{T}T_{R}^{1}+\overline{U}_{L}^{1}\rho _{1}^{+}h_{\rho 1j}^{D}D_{R}^{j}+%
\overline{T}_{L}^{1}\rho _{3}^{+}h_{\rho 1j}^{D}D_{R}^{j}  \notag \\
&&+\overline{U}_{L}^{n}\chi _{2}^{+}h_{\chi nj}^{D}D_{R}^{j}+\overline{U}%
_{L}^{1}\rho _{1}^{+}h_{\rho 1m}^{J}J_{R}^{m}+\overline{T}_{L}^{1}\rho
_{3}^{+}h_{\rho 1m}^{J}J_{R}^{m}+\overline{U}_{L}^{n}\chi _{2}^{+}h_{\chi
nm}^{J}J_{R}^{m}+h.c. \notag \\
\end{eqnarray}

From Eq. (\ref{lqm}) it follows that the mass matrices for the up and down
type quarks are given by: 
\begin{eqnarray}
M^{U}&=&\left( 
\begin{array}{ccccc}
v_{\eta }h_{\eta 11}^{U} & v_{\eta }h_{\eta 12}^{U} &  v_{\eta }h_{\eta
13}^{U} &\left| \right. & 0 \\ 
0 & 0 & 0 & \left| \right. & v_{\rho }h_{\rho 21}^{T} \\ 
0 & 0 & 0 & \left| \right. & v_{\rho }h_{\rho 31}^{T} \\ 
\text{\textemdash \hspace{0.1cm} \textemdash}&\text{\textemdash \hspace{0.1cm} \textemdash}&\text{\textemdash \hspace{0.1cm} \textemdash}&\text{\textemdash} &\text{\textemdash \hspace{0.1cm} \textemdash}\\
0 & 0 & 0 &  \left| \right. & v_{\chi }h_{\chi 11}^{T}%
\end{array}%
\right) = \left( 
\begin{array}{ccc}
M_U^{(A)} & \left| \right. &k_{3 \times 1} \\
\text{\textemdash \hspace{0.1cm} \textemdash} & \text{\textemdash}  &  \text{\textemdash  \hspace{0.1cm} \textemdash} \\ 
0_{1 \times 3} & \left| \right. & M_T%
\end{array}%
\right) \notag \\ \notag \\
M^{D}&=&\left( 
\begin{array}{cccccc}
v_{\rho }h_{\rho 11}^{D} & v_{\rho }h_{\rho 12}^{D} & v_{\rho }h_{\rho
13}^{D} & \left| \right. & v_{\rho }h_{\rho 12}^{J} & v_{\rho }h_{\rho 13}^{J} \\ 
0 & 0 & 0 &  \left| \right. & 0 & 0 \\ 
0 & 0 & 0 &  \left| \right. & 0 & 0 \\
\text{\textemdash \hspace{0.1cm} \textemdash}  & \text{\textemdash \hspace{0.1cm} \textemdash}  & \text{\textemdash \hspace{0.1cm} \textemdash}  &  \left| \right. & \text{\textemdash \hspace{0.1cm} \textemdash}  & \text{\textemdash \hspace{0.1cm} \textemdash}  \\ 
v_{\chi }h_{\chi 21}^{D} & v_{\chi }h_{\chi 22}^{D} & v_{\chi }h_{\chi
23}^{D} &  \left| \right. & v_{\chi }h_{\chi 22}^{J} & v_{\chi }h_{\chi 23}^{J} \\ 
v_{\chi }h_{\chi 31}^{D} & v_{\chi }h_{\chi 32}^{D} & v_{\chi }h_{\chi
33}^{D} &  \left| \right. & v_{\chi }h_{\chi 32}^{J} & v_{\chi }h_{\chi 33}^{J}%
\end{array}%
\right) = \left( 
\begin{array}{ccc}
M_D^{(B)} & \left| \right. &s_{3 \times 2} \\
\text{\textemdash \hspace{0.1cm} \textemdash} & \text{\textemdash}  &  \text{\textemdash  \hspace{0.1cm} \textemdash} \\ 
S_{2 \times 3} & \left| \right. & M_J%
\end{array}%
\right),\notag \\
\label{mixing-mass}
\end{eqnarray}%
where the diagonal blocks $M_{U}^{(A)}$ and $M_{D}^{(B)}$  are the same as  (\ref{U-textures}) and (\ref{D-textures}), $M_{T,J}$ are the masses of the $T^1$ and $J^n$ quarks, and $k, s$ and $S$ are mixing mass blocks. The different VEVs of the scalars have the following hierarchy: 
\begin{equation}
v_{\chi }>>v_{\rho },v_{\eta }\sim 246GeV.
\end{equation}

\subsection{Up sector}

The mass matrix for the up type quarks satisfies the following relation: 
\begin{equation}
M^{U}\left( M^{U}\right) ^{\dagger }=\left( 
\begin{array}{cc}
m_{t}^{2} & 0_{1\times 3} \\ 
0_{3\times 1} & \widetilde{M}%
\end{array}%
\right) ,\hspace{2cm}m_{t}^{2}=v_{\eta }^{2}\sum_{i=1}^{3}\left\vert h_{\eta
1i}^{U}\right\vert ^{2}
\end{equation}%
where $\widetilde{M}$ is given by: 
\begin{eqnarray}
\widetilde{M}&=&\left( 
\begin{array}{cccc}
v_{\rho }^{2}\left\vert h_{\rho 21}^{T}\right\vert ^{2} & v_{\rho
}^{2}h_{\rho 21}^{T}\left( h_{\rho 31}^{T}\right) ^{\ast } & \left| \right. & v_{\rho
}v_{\chi }h_{\rho 21}^{T}\left( h_{\chi 11}^{T}\right) ^{\ast } \\ 
v_{\rho }^{2}h_{\rho 31}^{T}\left( h_{\rho 21}^{T}\right) ^{\ast } & v_{\rho
}^{2}\left\vert h_{\rho 31}^{T}\right\vert ^{2} &\left| \right. & v_{\rho }v_{\chi }h_{\rho
31}^{T}\left( h_{\chi 11}^{T}\right) ^{\ast } \\ 
\text{\textendash \hspace{0.1cm} \textemdash \hspace{0.1cm}  \textemdash \hspace{0.1cm} \textemdash  \hspace{0.1cm} \textemdash}  & \text{ \textendash \hspace{0.1cm} \textemdash \hspace{0.1cm} \textemdash \hspace{0.1cm} \textemdash \hspace{0.1cm} \textemdash}  & \left| \right. & \text{\textendash \hspace{0.1cm}  \textemdash \hspace{0.1cm}  \textemdash \hspace{0.1cm}\textemdash \hspace{0.1cm} \textemdash}  \\
v_{\rho }v_{\chi }h_{\chi 11}^{T}\left( h_{\rho 21}^{T}\right) ^{\ast } & 
v_{\rho }v_{\chi }h_{\chi 11}^{T}\left( h_{\rho 31}^{T}\right) ^{\ast } & \left| \right. & 
v_{\chi }^{2}\left\vert h_{\chi 11}^{T}\right\vert ^{2}%
\end{array}%
\right) \notag \\
&=&\left( 
\begin{array}{ccc}
h_{\rho n1}^{T}\left( h_{\rho m1}^{T}\right) ^{\dagger }v_{\rho }^{2} & \left| \right.& 
h_{\rho n1}^{T}\left( h_{\chi 11}^{T}\right) ^{\ast }v_{\rho }v_{\chi } \\ 
\text{\textendash \hspace{0.1cm} \textemdash \hspace{0.1cm}  \textemdash \hspace{0.1cm} \textemdash  \hspace{0.1cm} \textemdash} & \left| \right. & \text{\textendash \hspace{0.1cm} \textemdash \hspace{0.1cm}  \textemdash \hspace{0.1cm} \textemdash  \hspace{0.1cm} \textemdash} \\
h_{\chi 11}^{T}\left( h_{\rho m1}^{T}\right) ^{\dagger }v_{\rho }v_{\chi } & \left| \right. & 
\left\vert h_{\chi 11}^{T}\right\vert ^{2}v_{\chi }^{2}%
\end{array}%
\right) .
\end{eqnarray}%
The submatrix $\widetilde{M}$ satisfies the following relation: 
\begin{equation}
\det \left( \widetilde{M}\right) =0.
\end{equation}%
Therefore, one quark (the $u$-quark) remains massless at tree level, one quark (the $c$-quark) will acquire mass through the mixing with the exotic $T$- quark, and two quarks (the $t$- and $T$- quarks) have tree-level masses without mixing. 
To find the rotation matrix
which diagonalizes the matrix $\widetilde{M}$, we perform a perturbative
diagonalization. The mass matrix $\widetilde{M}$ can be
block-diagonalized through the rotation matrix $W_{L}$, according to: 
\begin{equation}
W_{L}^{\dagger }\widetilde{M}W_{L}\simeq \left( 
\begin{array}{cc}
\widetilde{f} & 0_{2\times 1} \\ 
0_{1\times 2} & m_{T}^{2}%
\end{array}%
\right) ,
\end{equation}
with

\begin{equation}
W_{L}=\left( 
\begin{array}{cc}
1_{2\times 2} & B \\ 
-B^{\dagger } & 1%
\end{array}%
\right) ,\hspace{1cm}m_{T}^{2}\simeq \left\vert h_{\chi 11}^{T}\right\vert
^{2}v_{\chi }^{2}  \label{a4}
\end{equation}%
where $\widetilde{f}_{nm}=-v_{\rho }^{2}h_{\rho n1}^{T}\left( h_{\rho
m1}^{T}\right) ^{\dagger }$ with $m,n=2,3$. From the condition of the
vanishing of the off-diagonal submatrices in the previous expression, we
obtain at leading order in $B$ the following relations: 
\begin{equation*}
aB+b-Bm_T^2=0,\hspace{1cm}B^{\dagger }a+b^{\dagger }-m_T^2B^{\dagger }=0,
\end{equation*}%
\newline
where $a$ and $b$ have the following components: 
\begin{equation}
a_{nm}=h_{\rho n1}^{T}\left( h_{\rho m1}^{T}\right) ^{\dagger }v_{\rho }^{2},%
\hspace{1cm}b_{n1}=h_{\rho n1}^{T}\left( h_{\chi 11}^{T}\right) ^{\ast
}v_{\rho }v_{\chi }.
\end{equation}%
By using the method of recursive expansion taking into account the hierarchy 
$a_{nm}<<b_{n1}<<m_T^2$, we find that the submatrix $B$ is approximatelly given
by: 
\begin{equation}
B_{n1}\simeq \frac{v_{\rho }}{v_{\chi }}\frac{h_{\rho n1}^{T}}{h_{\chi
11}^{T}}\simeq \frac{m_{c}}{m_{T}}.
\end{equation}%
In sake of simplicity, let's assume that the Yukawa couplings $h_{\rho
m1}^{T}$ are real. In that case, the matrix $\widetilde{f}$ is diagonalized
by a rotation matrix: 
\begin{equation}
V_{Luc}=\frac{1}{\sqrt{\left( h_{\rho 11}^{T}\right) ^{2}+\left( h_{\rho
21}^{T}\right) }}\left( 
\begin{array}{cc}
h_{\rho 21}^{T} & -h_{\rho 31}^{T} \\ 
h_{\rho 31}^{T} & h_{\rho 11}^{T}%
\end{array}%
\right) ,
\end{equation}%
according to: 
\begin{equation}
V_{Luc}^{T}\widetilde{f}V_{Luc}=\widetilde{f}_{diag}=diag\left(
-m_{c}^{2},0\right) ,\hspace{0.5cm}m_{c}^{2}=v_{\rho }^{2}\left[ \left(
h_{\rho 21}^{T}\right) ^{2}+\left( h_{\rho 31}^{T}\right) ^{2}\right] ,%
\hspace{0.5cm}m_{u}=0.
\end{equation}
Since $v_{\rho }\sim v_{\eta }$, it follows that $\left\vert h_{\eta
1i}^{U}\right\vert >>\left\vert h_{\rho m1}^{T}\right\vert $. Here the following identity have been taking into account: 
\begin{equation}
\left( 
\begin{array}{cc}
\frac{c}{\sqrt{c^{2}+d^{2}}} & \frac{d}{\sqrt{c^{2}+d^{2}}} \\ 
-\frac{d}{\sqrt{c^{2}+d^{2}}} & \frac{c}{\sqrt{c^{2}+d^{2}}}%
\end{array}%
\right) \left( 
\begin{array}{cc}
c^{2} & cd \\ 
cd & d^{2}%
\end{array}%
\right) \left( 
\begin{array}{cc}
\frac{c}{\sqrt{c^{2}+d^{2}}} & -\frac{d}{\sqrt{c^{2}+d^{2}}} \\ 
\frac{d}{\sqrt{c^{2}+d^{2}}} & \frac{c}{\sqrt{c^{2}+d^{2}}}%
\end{array}%
\right) =\left( 
\begin{array}{cc}
c^{2}+d^{2} & 0 \\ 
0 & 0%
\end{array}%
\right) .
\end{equation}%
Then, it follows that the mass matrix $\widetilde{M}$ is diagonalized by a
rotation matrix $R_{L}$, according to: 
\begin{equation}
R_{L}^{T}\widetilde{M}R_{L}=diag\left( -m_{c}^{2},0,m_{T}^{2}\right) ,%
\hspace{1cm}\mathit{with}\hspace{1cm}R_{L}=\left( 
\begin{array}{cc}
V_{Luc} & B \\ 
-B^{\dagger }V_{Luc} & 1%
\end{array}%
\right) .
\end{equation}%
Therefore, the mass matrix $M^{U}\left( M^{U}\right) ^{\dagger }$ is
diagonalized by a rotation matrix $V_{L}^{U}$, according to: 
\begin{equation}
\left( V_{L}^{U}\right) ^{\dagger }M^{U}\left( M^{U}\right) ^{\dagger
}V_{L}^{U}=diag\left( m_{t}^{2},-m_{c}^{2},0,m_{T}^{2}\right) ,\hspace{1cm}%
\mathit{with}\hspace{1cm}V_{L}^{U}=\left( 
\begin{array}{cc}
1 & 0_{1\times 3} \\ 
0_{3\times 1} & R_{L}%
\end{array}%
\right) .
\end{equation}

\begin{figure}[tbh]
\includegraphics[width=20cm,height=25cm,angle=0]{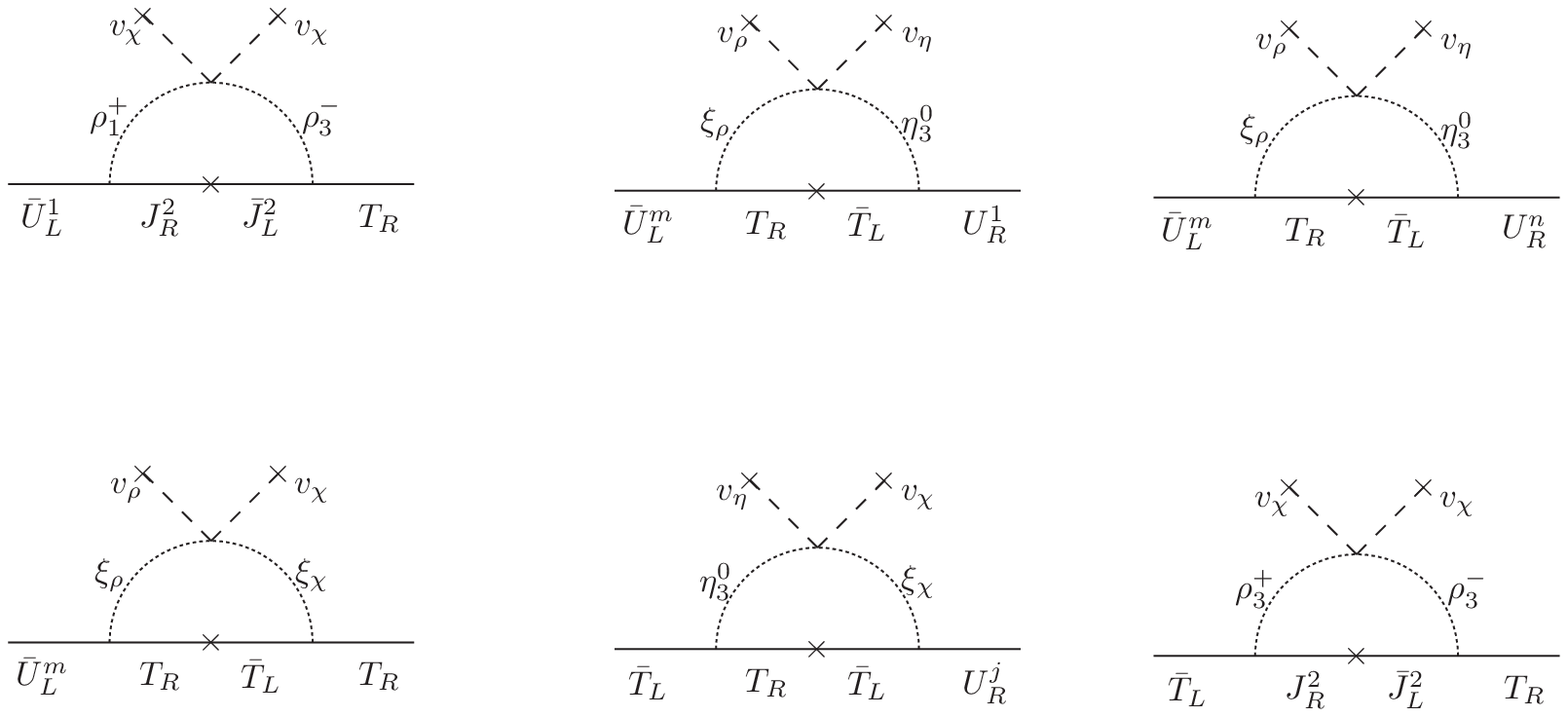}\vspace{-15cm}
\caption{One loop Feynman diagrams contributing to the entries of the up
type quark mass matrix.}
\label{figMu}
\end{figure}

Although the mixing terms with the exotic sector allow non vanishing mass to the $c$ quark, the lightest quark $u$ remains massless due to the zero-texture of $M^U$ as shown in Eq. (\ref{mixing-mass}) However, the vanishing entries can be filled by radiative corrections. In sake of simplicity we assume that the CP odd
neutral scalars are much heavier than the heavy exotic quarks $T$ and $J^{2}$%
, so that their loop contributions to the entries of the quark mass matrix
can be neglected. Here we do not consider the contributions coming from the
exotic quark $J^{3}$ since we assume that it does not mix with the SM quarks
and with the exotic quark $J^{2}$. Then, the heavy exotic quark $T$ with the
neutral scalars $\xi _{\rho }$, $\xi _{\chi }$, $\eta _{3}^{0}$ and the
heavy exotic quark $J^{2}$ with the charged scalars $\rho _{1}^{+}$ and $%
\rho _{3}^{\pm }$ running in the loop induce radiative corrections at one
loop level to most of the entries of the up type quark mass matrix thanks to
the scalar quartic interactions. These virtual scalars couple to real
neutral scalars which acquire VEVs after Electroweak Symmetry Breaking. In
this manner, the up quark mass is radiatively generated in an analogous way
to the loop induced neutrino mass generation processes. Besides that, we
assume that the quartic scalar couplings are approximatelly equal. Here we
also assume that $h_{\chi 11}^{T}>>\left\vert h_{\eta 1i}^{U}\right\vert $, $%
\left\vert h_{\rho m1}^{T}\right\vert $ and $h_{\chi mm}^{J}$ ($m=2,3$) is
much bigger than the magnitudes of the remaining down type quark Yukawa
couplings. These assumptions allow us to neglect the loop contributions to
the up and down type quark mass matrices that involve the mixings between
the SM quarks and the exotic quarks in the internal lines. Therefore, the leading one loop level contributions to the entries of the up type quark mass matrix come from the Feynman diagrams shown in Figure  \ref{figMu}. Here we use the Unitary gauge where we get rid of the Goldstone bosons $G^{\pm}_1$, $G^{\pm}_2$, $G^{0}_1$, $G^{0}_2$ and $G^{0}_3$. Hence, the
radiative corrections constraint the up type quark mass matrix to be of the
form:

\begin{equation}
M^{U}=\left( 
\begin{array}{cccc}
v_{\eta }h_{\eta 11}^{U} & v_{\eta }h_{\eta 12}^{U} & v_{\eta }h_{\eta
13}^{U} & \left( \delta M^{U}\right) _{14} \\ 
\left( \delta M^{U}\right) _{21} & \left( \delta M^{U}\right) _{22} & \left(
\delta M^{U}\right) _{23} & v_{\rho }h_{\rho 21}^{T}+\left( \delta
M^{U}\right) _{24} \\ 
\left( \delta M^{U}\right) _{31} & \left( \delta M^{U}\right) _{32} & \left(
\delta M^{U}\right) _{33} & v_{\rho }h_{\rho 31}^{T}+\left( \delta
M^{U}\right) _{34} \\ 
\left( \delta M^{U}\right) _{41} & \left( \delta M^{U}\right) _{42} & \left(
\delta M^{U}\right) _{43} & v_{\chi }h_{\chi 11}^{T}+\left( \delta
M^{U}\right) _{44}%
\end{array}%
\right) ,
\end{equation}

where their dominant loop induced entries are given by:

\begin{eqnarray}
\left( \delta M^{U}\right) _{14}&\simeq & -\frac{1}{16\pi ^{2}}\frac{\lambda
h_{\rho 12}^{J}h_{\rho 21}^{T}v_{\chi }^{2}}{m_{J_{2}}}C_{0}\left( \frac{%
m_{\rho _{1}^{+}}}{m_{J_{2}}},\frac{m_{\rho _{3}^{-}}}{m_{J_{2}}}\right) , \nonumber \\%
\left( \delta M^{U}\right) _{m1}&\simeq & -\frac{1}{16\pi ^{2}}%
\frac{\lambda h_{\rho m1}^{T}h_{\eta 11}^{U}v_{\eta }v_{\rho }}{m_{T}}%
C_{0}\left( \frac{m_{\xi _{\rho }}}{m_{T}},\frac{m_{\eta _{3}^{0}}}{m_{T}}%
\right) ,
\end{eqnarray}

\begin{eqnarray}
\left( \delta M^{U}\right) _{mn}&\simeq & -\frac{1}{16\pi ^{2}}\frac{\lambda
h_{\rho m1}^{T}h_{\eta 1n}^{U}v_{\eta }v_{\rho }}{m_{T}}C_{0}\left( \frac{%
m_{\xi _{\rho }}}{m_{T}},\frac{m_{\eta _{3}^{0}}}{m_{T}},\right) , \nonumber \\
\left( \delta M^{U}\right) _{m4}&\simeq & -\frac{1}{16\pi ^{2}}\frac{%
\lambda h_{\rho m1}^{T}h_{\chi 11}^{T}v_{\rho }v_{\chi }}{m_{T}}C_{0}\left( 
\frac{m_{\xi _{\rho }}}{m_{T}},\frac{m_{\xi _{\chi }}}{m_{T}}\right) ,
\end{eqnarray}

\begin{eqnarray}
\left( \delta M^{U}\right) _{4j}&\simeq & -\frac{1}{16\pi ^{2}}\frac{\lambda
h_{\chi 11}^{T}h_{\eta 1j}^{U}v_{\eta }v_{\chi }}{m_{T}}C_{0}\left( \frac{%
m_{\eta _{3}^{0}}}{m_{T}},\frac{m_{\xi _{\chi }}}{m_{T}}\right) , \nonumber \\
\left( \delta M^{U}\right) _{44}&\simeq & -\frac{1}{16\pi ^{2}}\frac{%
h_{\rho 12}^{J}h_{\rho 21}^{T}v_{\chi }^{2}}{m_{J_{2}}}D_{0}\left( \frac{%
m_{\rho _{3}^{\pm }}}{m_{J_{2}}}\right) ,
\end{eqnarray}
with $m,n=2,3$ and $j=1,2,3$. In the above equations we use the symbol $\lambda$ to indicate the quartic coupling terms from the scalar potential in (\ref{v00}), and the following functions have been
introduced:

\begin{eqnarray}
C_{0}\left( \widehat{m}_{1},\widehat{m}_{2}\right) &=&\frac{1}{\left( 1-%
\widehat{m}_{1}^{2}\right) \left( 1-\widehat{m}_{2}^{2}\right) \left( 
\widehat{m}_{1}^{2}-\widehat{m}_{2}^{2}\right) }\left\{ \widehat{m}_{1}^{2}%
\widehat{m}_{2}^{2}\ln \left( \frac{\widehat{m}_{1}^{2}}{\widehat{m}_{2}^{2}}%
\right) -\widehat{m}_{1}^{2}\ln \widehat{m}_{1}^{2}+\widehat{m}_{2}^{2}\ln 
\widehat{m}_{2}^{2}\right\} \nonumber \\ \nonumber \\
D_{0}\left( \widehat{m}_{1}\right) &=&\lim_{m_{2}\rightarrow m_{1}}C_{0}\left( 
\widehat{m}_{1},\widehat{m}_{2}\right) =\frac{-1+\widehat{m}_{1}^{2}-\ln 
\widehat{m}_{1}^{2}}{\left( 1-\widehat{m}_{1}^{2}\right) ^{2}}.
\end{eqnarray}

By assuming $m_{T}\sim m_{J^{2}}\sim v_{\chi }$ and $m_{\xi _{\rho }}\sim
m_{\xi _{\chi }}\sim m_{\eta _{3}^{0}}\sim m_{\rho _{1}^{\pm }}\sim m_{\rho
_{3}^{\pm }}$, it follows that the most important one loop correction for
vanishing entries of the tree level up type quark mass matrix is $\left(
\delta M^{U}\right) _{14}$. On the other hand, the one loop corrections of
the non vanishing entries of the tree level up type quark mass matrix can be
neglected when compared to their tree level values. Therefore, the dominant
one loop level contribution to $Tr\left( M^{U}\left( M^{U}\right) ^{\dagger
}\right) $ is roughly $\left\vert \left( \delta M^{U}\right)
_{14}\right\vert ^{2}$. Hence, the mass of the up quark can be estimated as: 
\begin{equation}
m_{u}\simeq \frac{1}{16\pi ^{2}}\frac{\lambda \left\vert h_{\rho
12}^{J}h_{\rho 21}^{T}\right\vert v_{\chi }^{2}}{m_{J_{2}}}C_{0}\left( \frac{%
m_{\rho _{1}^{+}}}{m_{J_{2}}},\frac{m_{\rho _{3}^{-}}}{m_{J_{2}}}\right) .
\end{equation}

Therefore, the smallness of the up quark mass can be explained by the loop
suppressed radiative seewaw-like process which involves a heavy exotic quark 
$J^{2}$ as well as virtual charged scalars $\rho _{1}^{+}$ and $\rho _{3}^{-}
$ whose corresponding Yukawa couplings have to be sufficiently small.

\subsection{Down sector}

The mass matrix for the down type quarks in (\ref{mixing-mass}) satisfies the following relation: 
\begin{equation}
M^{D}\left( M^{D}\right) ^{\dagger }=\left( 
\begin{array}{ccc}
c & 0_{1\times 2} & X_{1n} \\ 
0_{2\times 1} & 0_{2\times 2} & 0_{2\times 1} \\ 
X_{1n} ^{\dagger }& 0_{1\times 2} & Y_{nm}%
\end{array}%
\right) .
\end{equation}
with $n,m=2,3$, where

\begin{eqnarray}
c&=&\left[ \sum_{i=1}^{3}\left\vert h_{\rho 1i}^{D}\right\vert
^{2}+\sum_{n=2}^{3}\left\vert h_{\rho 1n}^{J}\right\vert ^{2}\right] v_{\rho
}^{2}, \nonumber \\
X_{1n}&=&\left[ \sum_{i=1}^{3}h_{\rho 1i}^{D}\left( h_{\chi ni}^{D}\right)
^{\dagger }+\sum_{m=2}^{3}h_{\rho 1m}^{J}\left( h_{\chi nm}^{J}\right)
^{\dagger }\right] v_{\rho }v_{\chi },\nonumber \\
Y_{nm}&=&\left[
\sum_{i=1}^{3}h_{\chi ni}^{D}\left( h_{\chi mi}^{D}\right) ^{\dagger
}+\sum_{p=2}^{3}h_{\chi np}^{J}\left( h_{\chi mp}^{J}\right) ^{\dagger }%
\right] v_{\chi }^{2},
\label{down-components}
\end{eqnarray}
from which it follows that:

\begin{equation}
\det \left[ M^{D}\left( M^{D}\right) ^{\dagger }\right] =0.
\end{equation}

Therefore, the mass matrix texture $M^{D}$ leads to a massless down and
strange quarks, which is not phenomenological viable. Besides that, the mass
matrix $M^{D}\left( M^{D}\right) ^{\dagger }$ is partially diagonalized by a
rotation matrix $V_{L}^{D}$, according to: 
\begin{equation}
\left( V_{L}^{D}\right) ^{\dagger }M^{D}\left( M^{D}\right) ^{\dagger
}V_{L}^{D}\simeq \left( 
\begin{array}{cccc}
m_{b}^{2} & 0 & 0 & 0_{1\times 2} \\ 
0 & 0 & 0 & 0_{1\times 2} \\ 
0 & 0 & 0 & 0_{1\times 2} \\ 
0_{2\times 1} & 0_{2\times 1} & 0_{2\times 1} & Y%
\end{array}%
\right) ,
\end{equation}
where

\begin{equation}
V_{L}^{D}=\left( 
\begin{array}{ccc}
1 & 0_{1\times 2} & F \\ 
0_{2\times 1} & 1_{2\times 2} & 0_{2\times 1} \\ 
-F^{\dagger } & 0_{1\times 2} & 1_{2\times 2}%
\end{array}%
\right)
\end{equation}
and%

\begin{equation}
m_{b}^{2}\simeq c-2\sum_{n,m=2}^{3}X_{1n}Y_{nm}^{-1}X_{1m}^{\dagger },%
\hspace{1cm} F_{1m}=\sum_{n=2}^{3}X_{1n}Y_{nm}^{-1}.
\end{equation}
This shows that radiative corrections at one loop level have to be
introduced in order to generate the masses for the down and strange quarks. In sake of simplicity, we assume a diagonal base for the exotic quarks $J_2$ and $J_3$, and the hierarchy $M_{J_{3}} \gg M_{J_{2}}$, which suppress the mixing terms with the $J_3$ quark at low energy.     
These assumptions implies that the masses of the exotic quarks $J^{2}$ and $J^{3}$ are
given by:

\begin{equation}
M_{J_{2}}=v_{\chi }h_{\chi 22}^{J},\hspace{2cm}M_{J_{3}}=v_{\chi }h_{\chi
33}^{J}.
\end{equation}

\begin{figure}[tbh]
\vspace{-5cm} \includegraphics[width=20cm,height=25cm,angle=0]{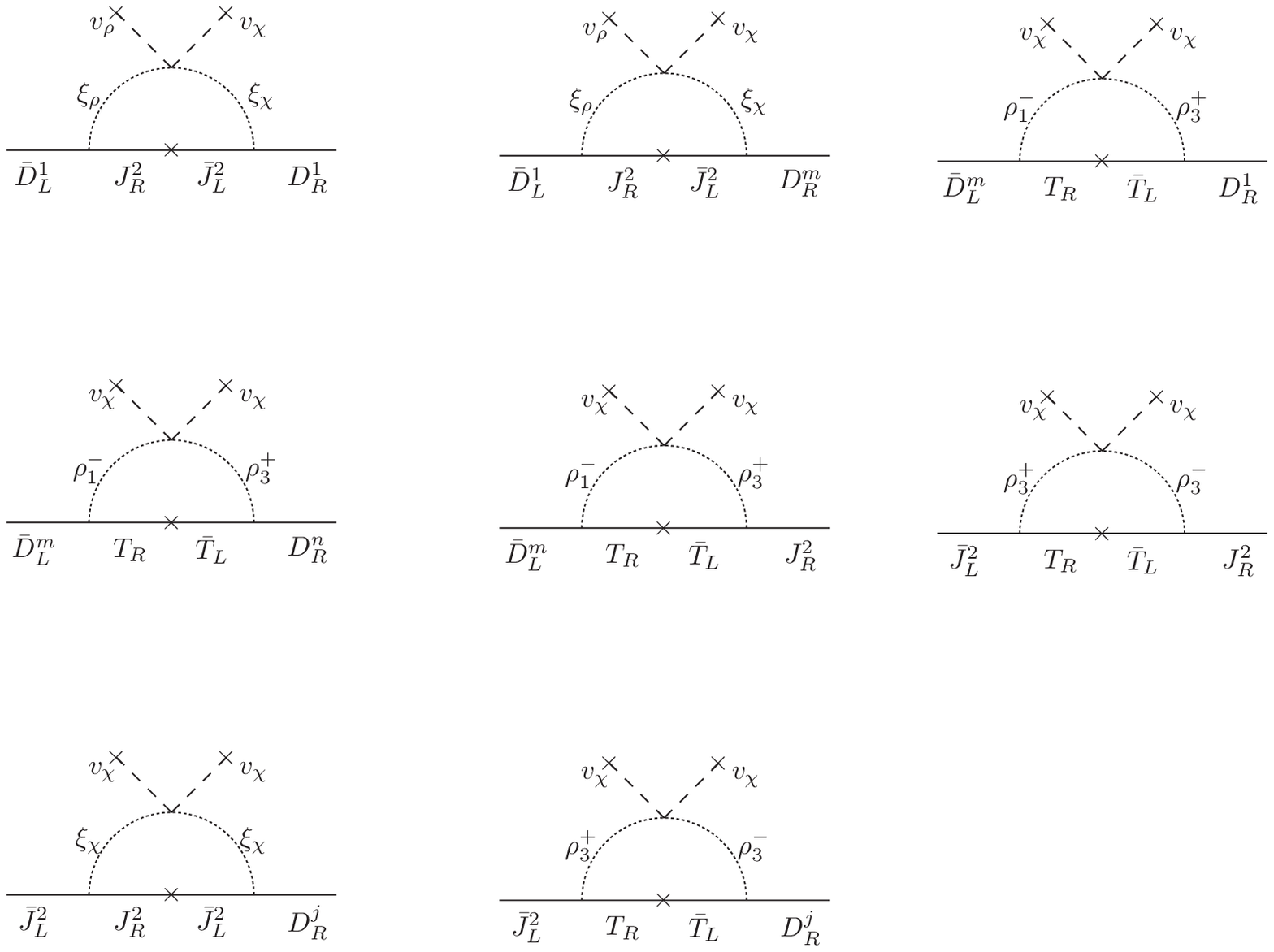}
\vspace{-3cm}
\caption{One loop Feynman diagrams contributing to the entries of the down
type quark mass matrix.}
\label{figMd}
\end{figure}

Some entries of the down type quark mass matrix receive loop corrections
involving neutral scalars $\xi _{\rho }$, $\xi _{\chi }$ with the heavy
exotic quark $J^{2}$ and charged scalars $\rho _{1}^{\pm }\ $and $\rho
_{3}^{\pm }$\ with heavy exotic quark $T$ running in the internal lines of
the loops. These virtual scalars couple to real neutral scalars due to the
scalar quartic interactions. The leading one loop level contributions to the
entries of the down type quark mass matrix come from the Feynman diagrams
shown in Figure \ref{figMd}. After these radiative corrections are taken
into account, the down type quark mass matrix takes the following form:

\begin{equation}
M^{D}=\left( 
\begin{array}{ccccc}
v_{\rho }h_{\rho 11}^{D}+\left( \delta M^{D}\right) _{11} & v_{\rho }h_{\rho
12}^{D}+\left( \delta M^{D}\right) _{12} & v_{\rho }h_{\rho 13}^{D}+\left(
\delta M^{D}\right) _{13} & v_{\rho }h_{\rho 12}^{J} & 0 \\ 
\left( \delta M^{D}\right) _{22} & \left( \delta M^{D}\right) _{23} & 0 & 
\left( \delta M^{D}\right) _{24} & 0 \\ 
\left( \delta M^{D}\right) _{32} & \left( \delta M^{D}\right) _{33} & 0 & 
\left( \delta M^{D}\right) _{34} & 0 \\ 
v_{\chi }h_{\chi 21}^{D}+\left( \delta M^{D}\right) _{41} & v_{\chi }h_{\chi
22}^{D}+\left( \delta M^{D}\right) _{42} & v_{\chi }h_{\chi 23}^{D}+\left(
\delta M^{D}\right) _{43} & v_{\chi }h_{\chi 22}^{J}+\left( \delta
M^{D}\right) _{44} & 0 \\ 
0 & 0 & 0 & 0 & v_{\chi }h_{\chi 33}^{J}%
\end{array}%
\right)
\end{equation}
where their dominant loop corrections are given by:

\begin{eqnarray}
\left( \delta M^{D}\right) _{11}&\simeq& -\frac{1}{16\pi ^{2}}\frac{\lambda
h_{\rho 12}^{J}h_{\chi 21}^{D}v_{\rho }v_{\chi }}{m_{J_{2}}}C_{0}\left( 
\frac{m_{\xi _{\rho }}}{m_{J_{2}}},\frac{m_{\xi _{\chi }}}{m_{J_{2}}}\right), \nonumber \\
\left( \delta M^{D}\right) _{1m}&\simeq& -\frac{1}{16\pi ^{2}}%
\frac{\lambda h_{\rho 12}^{J}h_{\chi 2m}^{D}v_{\rho }v_{\chi }}{m_{J_{2}}}%
C_{0}\left( \frac{m_{\xi _{\rho }}}{m_{J_{2}}},\frac{m_{\xi _{\chi }}}{%
m_{J_{2}}}\right) ,
\end{eqnarray}

\begin{eqnarray}
\left( \delta M^{D}\right) _{m1}&\simeq & -\frac{1}{16\pi ^{2}}\frac{\lambda
h_{\rho m1}^{T}h_{\rho 11}^{D}v_{\chi }^{2}}{m_{T}}C_{0}\left( \frac{m_{\rho
_{1}^{-}}}{m_{T}},\frac{m_{\rho _{3}^{+}}}{m_{T}}\right) ,\nonumber \\%
\left( \delta M^{D}\right) _{mn}&\simeq &-\frac{1}{16\pi ^{2}}\frac{\lambda
h_{\rho m1}^{T}h_{\rho 1n}^{D}v_{\chi }^{2}}{m_{T}}C_{0}\left( \frac{m_{\rho
_{1}^{-}}}{m_{T}},\frac{m_{\rho _{3}^{+}}}{m_{T}}\right) ,
\end{eqnarray}

\begin{eqnarray}
\left( \delta M^{D}\right) _{m4}&\simeq & -\frac{1}{16\pi ^{2}}\frac{\lambda
h_{\rho m1}^{T}h_{\rho 12}^{J}v_{\chi }^{2}}{m_{T}}C_{0}\left( \frac{m_{\rho
_{1}^{-}}}{m_{T}},\frac{m_{\rho _{3}^{+}}}{m_{T}}\right) ,\nonumber \\%
\left( \delta M^{D}\right) _{44}&\simeq & -\frac{1}{16\pi ^{2}}\frac{\lambda
h_{\rho 21}^{T}h_{\rho 12}^{J}v_{\chi }^{2}}{m_{T}}D_{0}\left( \frac{m_{\rho
_{3}^{\pm }}}{m_{T}}\right)
\end{eqnarray}

\begin{eqnarray}
\left( \delta M^{D}\right) _{4j} &\simeq & -\frac{1}{16\pi ^{2}}\left[ \frac{%
\lambda h_{\chi 22}^{J}h_{\chi 2j}^{D}v_{\chi }^{2}}{m_{J_{2}}}D_{0}\left( 
\frac{m_{\xi _{\chi }}}{m_{J_{2}}}\right) +\frac{\lambda h_{\rho
21}^{T}h_{\rho 1j}^{J}v_{\chi }^{2}}{m_{T}}D_{0}\left( \frac{m_{\rho
_{3}^{\pm }}}{m_{T}}\right) \right] ,
\end{eqnarray}
with $m,n=2,3$ and $j=1,2,3$. By assuming $m_{T}\sim m_{J^{2}}\sim v_{\chi }$ and $m_{\xi _{\rho }}\sim
m_{\xi _{\chi }}\sim m_{\rho _{1}^{\pm }}\sim m_{\rho _{3}^{\pm }}$, it
follows that the two most important loop corrections to $Tr\left(
M^{D}\left( M^{D}\right) ^{\dagger }\right) $ come from terms of the order $%
\sum_{m=2}^{3}\left\vert \left( \delta M^{D}\right) _{m4}\right\vert ^{2}$
and $\sum_{m=2}^{3}\sum_{n=2}^{3}\left\vert \left( \delta M^{D}\right)
_{mn}\right\vert ^{2}$. \ These terms come from the loop corrections of the
tree level vanishing entries of the down type quark mass matrix. On the
other hand, as in the up sector, the one loop corrections for the non
vanishing entries of the tree level down type quark mass matrix can be
neglected when compared to their tree values. Therefore, for the case $%
\left\vert h_{\rho 1n}^{D}\right\vert <<\left\vert h_{\rho
12}^{J}\right\vert $ with $n=1,2$, the masses of the down and strange quarks
can be estimated as:

\begin{eqnarray}
m_{d}&\simeq &\frac{1}{16\pi ^{2}}\frac{\lambda v_{\chi }^{2}}{m_{T}}\sqrt{%
\sum_{m=2}^{3}\sum_{n=2}^{3}\left\vert h_{\rho m1}^{T}h_{\rho
1n}^{D}\right\vert ^{2}}C_{0}\left( \frac{m_{\rho _{1}^{-}}}{m_{T}},\frac{%
m_{\rho _{3}^{+}}}{m_{T}}\right) ,\nonumber \\ 
m_{s} &\simeq &\frac{1}{16\pi
^{2}}\frac{\lambda \left\vert h_{\rho 12}^{J}\right\vert v_{\chi }^{2}}{m_{T}%
}\sqrt{\sum_{m=2}^{3}\left\vert h_{\rho m1}^{T}\right\vert ^{2}}C_{0}\left( 
\frac{m_{\rho _{1}^{-}}}{m_{T}},\frac{m_{\rho _{3}^{+}}}{m_{T}}\right) .
\end{eqnarray}

We can see that the charged scalar loop contributions are crucial to give
masses to the down and strange quarks. Besides that, the lightness of the
down quark can be explained from the smallness of $\left\vert h_{\rho
1n}^{D}\right\vert $ as well as from the loop suppressed radiative
seewaw-like process which involves a heavy exotic quark $T$ as well as
virtual charged scalars $\rho _{1}^{-}$ and $\rho _{3}^{+}$. Furthermore,
the inequality $\left\vert h_{\rho 1n}^{D}\right\vert <<\left\vert h_{\rho
12}^{J}\right\vert $ can explain the hierarchy between the down and strange
quark masses. 

\section{Conclusions}

In this paper we have discussed the generation of quark masses in a model
based on the gauge symmetry $SU(3)_{c}\otimes SU(3)_{L}\otimes U(1)_{X}$
where\ this symmetry is spontaneously broken to the SM electroweak group $%
SU(2)_{L}\otimes U(1)_{Y}$ at the TeV \ scale. The abelian non-universal $%
U(1)_{X}$ symmetry in the quark sector exhibited in this 3-3-1 model leads
to the tree level cancellation of the Yukawa couplings not allowed by the
symmetry.  Indeed, since the $U(1)_X$ symmetry of the model distinguishes one family from the other two, the zero-texture structures obtained by (\ref{U-textures}) and (\ref{D-textures}) arise naturally, which will lead that only one family (the third) obtain tree-level masses. The $U(1)_X$ quantum numbers for the exotic quarks $T$ and $J$ are obtained by the condition of cancellation of  anomalies, which leads to the mixing terms shown in the third line in Eq. (\ref{yukawaAB}). These mixing couplings will produce a tree-level mass for the middle quark (charm-quark), while the lighter quarks remain massless due to the symmetry. Thus, it is necessary to generate radiative corrections involving scalars and exotic quarks in the internal lines in order to obtain the complete mass spectrum. 
In this framework we assume that the CP odd neutral scalars are much heavier that the heavy
exotic quarks $T$ and $J^{2}$ and we restrict to the scenario characterized
by the absence of mixing between the heavy exotic quark $J^{3}$ and the
remaining down type quarks. We have found that the mixings between the SM
quarks and the exotic quarks as well as the seesaw-like radiative mechanism
are crucial to explain the hierarchy of the quark mass spectrum.

\section*{Acknowledgments}

R. Martinez and F. Ochoa thank to Colciencias for financial support

\newpage

\end{document}